# Electron-induced $CO_2$ and hydrocarbon sputtering of functionalized hydrocarbons in icy planetary analogs

Sankhabrata Chandra, Bryana L. Henderson*, Murthy S. Gudipati

Science Division, Jet Propulsion Laboratory, California Institute of Technology, Pasadena, CA, 91109, USA

*Corresponding author: Bryana L. Henderson (bryana.l.henderson@jpl.nasa.gov)

**Abstract**

$CO_2$ has been detected in both the tenuous exosphere and surface chaos regions of Europa, but it is still unclear whether this $CO_2$ is generated in situ by radiolysis or whether it is directly delivered by the ocean. In this work, we study the radiolysis pathway, and explore the possibility that organics upwelled from the subsurface oceans could be contributing to this signature on the surface and in the exosphere. Specifically, we report here on the evolution of carbon-containing byproducts generated by electron-induced sputtering of organics with different functional groups – hexanoic acid, hexanol, and hexane – in water ice. We found that, upon electron irradiation in vacuum, the acid-functionalized molecules generated a factor of over 10x more $CO_2$ than either the non-functionalized or alcohol-functionalized molecules, but that all three species produced $CO_2$ to some extent. The amount of $CO_2$ produced was found to depend upon temperature. $CO_2$ was the dominant product for ices at 100 K and 120 K, and production of $CO_2$ was 3x higher at 100 and 120 K than at 80 K. Sputtering of long chain molecules such as pentane was a factor of 100x higher in the hexanoic acid-containing ice than in the hexane/hexanol ice, suggesting that organics with carboxylic acid (COOH) functional groups may be also more likely to produce volatile species that can be ejected into the exosphere.

## 1. Introduction

Carbon is the fourth most abundant element in the universe, and out of the different carbon-containing species, $CO_2$ is frequently observed in different planetary bodies (Ahrens et al., 2022a; Colaprete et al., 2010; Howard, 2000; Phillips et al., 2011; Rubanenko et al., 2018). Galileo's NIMS data shows features of $CO_2$ ice infrared absorption band on Jupiter's moon Europa, Ganymede, and Callisto (Hansen and McCord, 2008; Hibbitts et al., 2003, 2002) most recently JWST observations confirmed Galileo data (Trumbo and Brown, 2023). The Cassini spacecraft has detected $CO_2$ at Enceladus (Combe et al., 2019) and organic-rich Titan (de Kok et al., 2007; Samuelson et al., 1983). $H_2O$-rich $CO_2$ ice has been observed on Hyperion, Iapetus's pole, Phoebe, and Dione (Clark et al., 2019; Cruikshank et al., 2010; Palmer and Brown, 2011), as well as on Uranian and Neptune systems (Cartwright et al., 2015; Grundy et al., 2010). In some cases, it is unclear whether the $CO_2$ is purely endogenic or whether exogenic delivery or radiation could have contributed significantly to the presence of $CO_2$ on these bodies (Ahrens et al., 2022b). Very recently, it was suggested that the origin of $CO_2$ on Europa may be primarily endogenic, as it is associated with young chaos terrains (Trumbo and Brown, 2023). However, the radiolysis of complex organic molecules (COMs) can also generate $CO_2$ along with other volatile byproducts (Gomis and Strazzulla, 2005; Raut et al., 2012). Various forms of COMs have been detected in the Solar System, including hydrocarbons at Ganymede and Callisto's surfaces (McCord et al., 1997), at Saturn's moon Enceladus (Postberg et al., 2018), and at Titan (Nixon, 2024). Oxygen-containing COMs, including alcohols, aldehydes, and acids have also been detected at Enceladus (Peter et al., 2024) and Titan (Nixon, 2024). A large variety of COMs, including aliphatic and aromatic hydrocarbons, alcohols, aldehydes and acids have been detected in the Jupiter-family comet 67P/Churyumov-Gerasimenko (Hänni et al., 2022) and are likely to be present in other comets. When exposed to radiation, these COMs may undergo chemical and physical effects such as radiolysis or sputtering, which in turn can influence a body's surface and exospheric compositions, and perhaps even generate local enhancements that could be probed by future missions. Understanding these effects is crucial to comprehending the exchange between icy surfaces and their exospheres (Johnson et al., 2004).

Spectroscopy has been used for several decades to track in-situ radiation-induced changes in COM-containing ices. Hydrocarbon and water-ice mixtures such as $CH_4$:$H_2O$ (Moore and Hudson, 1998; Wada et al., 2006), $C_2H_6$:$H_2O$ ice (Barros et al., 2016; Khare et al., 1993), $C_3H_8$:$H_2O$ ice (Behmard et al., 2019; Hand and Carlson, 2012) and $C_4H_{10}$:$H_2O$ (Hand and Carlson, 2012) ice have been studied using a variety of energetic particle sources. Following bombardment, these systems have been shown to generate alcohols, aldehydes, acids, $CO_2$ and CO in the ices. Electron irradiation experiments with three- and four-carbon-containing hydrocarbon mixtures with $H_2O$ ice at 70-100 K have suggested the destruction of the hydrocarbon compounds and in-situ spectroscopic detection of long chain, branched aliphatic, aldehydes, ketones, esters, alcohols and $CO_2$ (Hand and Carlson, 2012). Similarly, UV or electron or ion irradiation of methanol or formic-acid-containing ices have been shown to generate formaldehyde, methane, CO, $CO_2$, and formyl radicals (Allamandola et al., 1988; Baratta et al., 1994; Gerakines et al., 1996; Hudson and Moore, 2000; Palumbo et al., 1999). These products have mostly been detected in situ using infrared spectroscopy, or in some cases, by laser ablation ionization mass spectrometry, which excavates and detects the products in low temperature ice without warming up the sample (Henderson and Gudipati, 2015). Radiation products have also been monitored by temperature-programmed mass spectrometry during warm-up (Zhu et al., 2018), or detected in the residues remaining after warming, by a variety of other standard laboratory techniques (Couturier-Tamburelli et al., 2024). So far, these studies have primarily focused on the evolution of products inside the ice, rather than on the products sputtered away from the ice during irradiation. Our goal is to study how the functional groups of COMs influence their sputtering behavior over time and at different temperatures.

The majority of sputtering experiments to date have focused on water ice, which generates simple species such as H, $H_2$, O, $O_2$, OH and $H_2O$ (Baragiola et al., 2003; Bar-Nun et al., 1985; Davis et al., 2021; Galli et al., 2018; Teolis et al., 2005) with a yield that depends on temperature of the ice (Davis et al., 2021). Non-ice components are likely to participate in the radiation-induced chemistry on these icy bodies (Hansen and McCord, 2008; McCord et al., 1997) and increase the complexity of the sputtered distributions. To the best of our knowledge, correlation between the type and evolution of sputtered byproducts and the type of chemical functional groups initially present in cryogenic ices has not yet been systematically investigated. Therefore, in this work, we use water ice mixtures containing model organic compounds (hexane, hexanol, and hexanoic acid), which each contain a different type of chemical functional group, to assess the differences in the observed sputtering products, particularly focusing on production of $CO_2$. The evolution of the sputtered products are studied at 80 K, 100 K and 120 K since the average temperatures of many bodies lie within this range: 96 K for Europa (Ashkenazy, 2019), 93 K for Titan (Jennings et al., 2019), 75 K for Enceladus (Spencer et al., 2006) and 120 K for Ganymede and Callisto (Squyres, 1980). Therefore, sputtering at these temperatures will provide some insight into the exchange of materials between surface and exospheres of typical icy planetary bodies.

## 2. Experimental Setup

Electron sputtering experiments were performed at the Jet Propulsion Laboratory's Ice Spectroscopy Laboratory (ISL). The experimental setup combines an electron gun (Model: ELG-2/EGPS-1022, Kimball Physics, Inc., USA) and a quadrupole mass spectrometer (QMS) (RGA200, Stanford Research Systems, Inc., USA) mounted in a stainless-steel vacuum chamber (purchased from Kimball Physics, Inc., USA), shown in Figure 1. The vacuum chamber pressure is monitored by a hot cathode ionization gauge (Model: KJLC354401YF, Kurt J. Lesker Company, USA) which can operate from $6.6 \times 10^{-2}$ mbar to $1.3 \times 10^{-9}$ mbar and the vacuum is maintained by an 80 liter/second turbomolecular pump (Model: Twistorr 84FS, Agilent Technologies, USA) backed by a dry scroll pump. With this pumping system, the vacuum is maintained at or below $10^{-8}$ mbar at room temperature. All of the experiments are performed at 80 K, 100 K and 120 K and the vacuum chamber pressure during these experiments came down to the range of $1.0 \times 10^{-9}$ mbar. The temperature of the sample holder is maintained by a closed-cycle helium cryostat (Advanced Research Systems, Inc., USA). A 150 mm tall copper rod is attached to the cold head, and the electron gun is positioned at a 45-degree angle at a distance of 50 mm so that the electron beam fills the full width of the copper rod's sample well (20 mm in diameter). The temperature of the sample holder is monitored using two programmable cryogenic temperature sensors (DT-670 silicon diode, LakeShore Cryotronics, Inc., USA). These are placed at the bottom of the copper rod (Sensor A) and 15 mm from the top of the sample holder (Sensor B). For ice deposition, two different techniques are used depending upon the volatility of the sample: vapor deposition and spray deposition. For vapor deposition experiments, degassed ultrapure water (JT Baker Chemicals, Inc., USA) and hexane (HPLC grade, Fisher Scientific Co LLC) were introduced through two separate leak valves. Before vapor deposition, three freeze-pump-thaw cycles were performed at a pressure of $10^{-8}$ mbar to remove any unwanted dissolved gases in the liquids. $H_2O$:hexane ice is formed by co-deposition of $H_2O$ and hexane for 30 minutes at a combined pressure of $1 \times 10^{-6}$ mbar to generate a sample of a few microns thick, based on laser interference studies on ices made in our lab under similar conditions (Barnett et al., 2012).

For preparing $H_2O$:hexanoic acid ice and $H_2O$:hexanol ice, the vapor deposition technique was not suitable because of the lower volatilities of the hexanoic acid (boiling point of 476 K) and hexanol (boiling point of 430 K). Therefore, a spray technique was adopted. Before spraying, the sample holder was first cooled down to 30 K. A temperature profile of the two temperature diodes during the experiment is plotted in Figure A1 in Appendix. When the sample holder temperature reached ~30 K, as confirmed by the uppermost Sensor B in inset plot of Figure A1, it was separated from the cold head by a wobble stick and was brought into the small spray deposition antechamber on the right-hand side as shown in Figure 1. Temperature drops can be observed for sensors A and B due to removal of the sample holder from the cold head. A gate valve was then closed to isolate the antechamber (see Figure 1), where a scroll pump maintained the vacuum at about $10^{-2}$ mbar. A homemade nebulizer was then used to spray 100±10 microliters of $H_2O$:hexanoic acid or $H_2O$:hexanol onto the cold copper substrate. The nebulizer's design includes nested stainless-steel tubes, where a syringe and 300 mm needle were used to inject the solution into the inner tube and a dry carrier gas (here, dry nitrogen) was used to make the droplets. During this spray the base pressure in the side chamber was maintained at ~550 mbar. The approximate duration of spraying was about one minute. Aqueous solutions of 1% hexanoic acid or hexanol were used for these depositions, since this is close to the maximum solubility of hexanoic acid and hexanol in water. After completing the spray, the carrier gas valve was closed (see inset in Figure 1), and the antechamber containing the sample ice was pumped down for 10 minutes to $<10^{-7}$ mbar pressure with the help of the turbomolecular pump (which was temporarily isolated from the main chamber with a valve during this time). The turbomolecular pump valve and chamber gate valve were then reopened, and the sample was slid back into position onto the copper cold head. Upon reattachment to the cold head, the temperature measurements for both Sensor A and Sensor B rose a few K which can be seen as sharp peaks in the traces in Figure A1. After removing the wobble stick, closing the gate valve, and re-equilibrating the sample, the substrate was ramped to the desired temperature (80 K, 100 K or 120 K). For each sputtering experiment, an electron energy of 2 keV is used with an electron current of 5 microamp for 120 minutes. The focus and grid voltages were maintained at 0.2 V and 0.80 V, respectively. After considering the geometrical arrangement of the electron gun, electron energy, electron current, and the voltages of focus and grid, the total energy flux on the ice surface is about $1.1 \times 10^{13}$ keV cm$^{-2}$ second$^{-1}$. Our 120 minutes of radiation is equivalent to approximately 360 hours on Europa, $8.5 \times 10^4$ hours on Ganymede, $10^5$ hours on Callisto, and $2.7 \times 10^4$ hours on Enceladus (Johnson et al., 2004; Paranicas et al., 2012). However, it should be noted that this calculation does not include irradiation by protons or other ions that are likely to be bombarding the planetary surfaces.

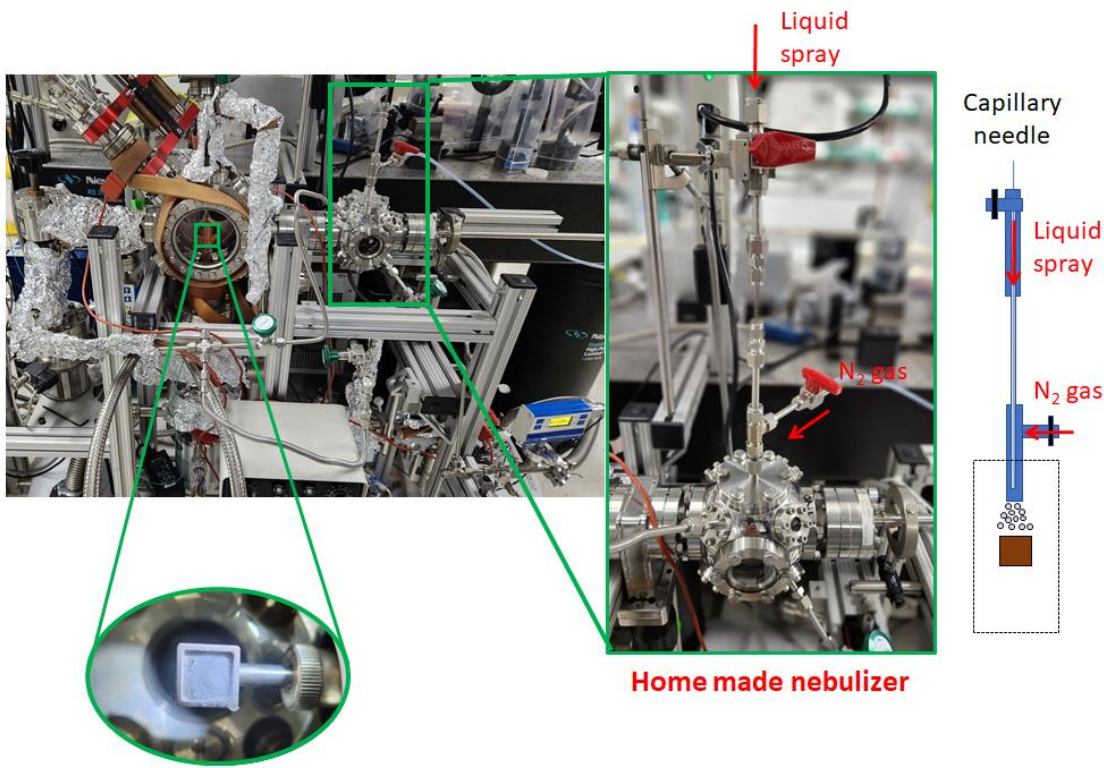

Figure 1: Experimental setup for the electron sputtering experiments, installed at the Jet Propulsion Laboratory's Ice Spectroscopy Laboratory (ISL). Left picture shows the full experimental chamber with two close-up views of ice formation (left) and the home-made nebulizer (right).

## 3. Results and Discussion

### 3.1 Control experiments

### 3.1.1 Deposition method controls

A major goal of this work was to evaluate the differences in sputtered $CO_2$ abundance during irradiation of organics with different functional groups, so we first conducted a series of control experiments to evaluate the possibility of any confounding contributions to m/z 44. Since all the masses are detected in the QMS, the detected masses are represented as cations throughout all experiments. In our system, $CO_2$ appears at m/z= 44 and also as a doubly-charged species at a mass to charge ratio of 22. These two masses are plotted in Figure A2 for a $H_2O$:hexanoic acid experiment at 120 K, where it can be seen that, throughout the duration of the experiment, m/z=22 follows the same trend as the m/z=44 peak. The ratio of m/z=22 to m/z=44 is approximately 1%, which is similar to the percentage given in the NIST electron ionization mass spectrometry library for the mass spectrum of $CO_2$ (E. Wallace, n.d.) (see Figure A2). The sputtered intensity of nearby masses m/z=21 and 23 in Figure A2 show the baseline to be one order of magnitude lower than m/z=22. Therefore, we assigned m/z=44 in all experiments as $CO_2$ and presume here that the contribution from other m/z=44 species is minimal.

For evaluating the $CO_2$ sputtering, it was also necessary to quantify how much ambient $CO_2$ (which is naturally present in the chamber and can be drawn in during sample injection) is trapped during spray deposition of the sample ice. Figure 2 compares the $CO_2$ sputtered when irradiated with 2 keV electrons for 1) our vapor deposition technique (which uses degassed water) and 2) our spray deposition technique (which uses non-degassed water). The amount of $CO_2$ from these two techniques are compared with the amount that is collected during a mock spray deposition (injection of ambient air). The results show that all three of these control experiments gave a similar amount of $CO_2$, with the integrated amount varying by less than 10% between techniques.

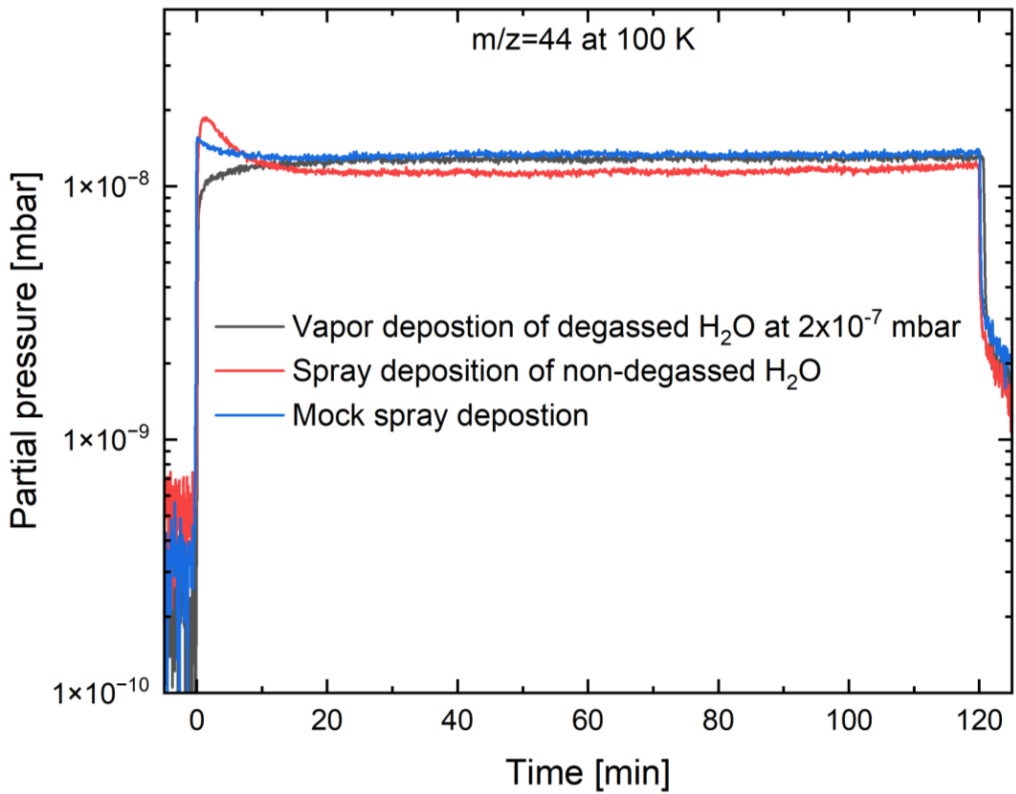

Figure 2: Tracking of $CO_2$ (m/z=44) during sputtering of three control samples: no water deposited (a "mock" spray deposition of air onto a bare substrate, blue) vs. pure water ice generated by vapor deposition (black) vs. pure water ice generated by spray deposition (red). All three control experiments generate approximately the same amount of $CO_2$, suggesting that a small amount is present in the chamber as a residual gas that is incorporated at similar levels in all three experiments.

To perform the $H_2O$:hexanoic acid and $H_2O$:hexanol sputtering experiments, 100 microliters of the stock solution is sprayed through the home-built nebulizer onto the precooled sample holder block and the sample is transferred into the electron sputtering chamber, as described earlier. The electron sputtering experiment is performed for 120 minutes with an energy of 2 keV and 5 μA current at the source. Figure A3 shows the sputtering evolution of m/z 44 in the $H_2O$ ice, $H_2O$:hexanol ice and $H_2O$:hexanoic acid ice at 100 K temperature (all spray deposited). $H_2O$:hexanoic acid ice sputtered 78% more $CO_2$ and $H_2O$:hexanol ice sputtered 24% more $CO_2$ at 100 K in comparison to the pure water ice experiment. In each experiment we have subtracted either the spray-deposited or vapor-deposited "background $CO_2$" (in each case corresponding to the sample's deposition technique and temperature) so that the $CO_2$ production from different functionalized organics could be more accurately compared.

### 3.1.2 Repeatability of the spray technique

To determine the repeatability of our spray technique, three identical hexanoic acid experiments have been performed at 100 K and $CO_2$ yields from these three experiments are plotted in Figure A4 in the Appendix. This shows that at the very beginning of the sputtering, the $CO_2$ concentration rises reproducibly by three orders of magnitude to $1.5 \times 10^{-7}$ mbar, whereas for pure water ice the $CO_2$ sputtering remains in the range of 1 x $10^{-8}$ mbar. The mean and standard deviation is also shown in Figure A4 inset for three experiments and the relative standard deviation is in the range of ~30% for the first 10 minutes and then it reduces to ~3% afterward. The $CO_2$ yield decays exponentially during the hexanoic acid sputtering experiment as the -COOH fragments from the hydrocarbon chain due to the electron sputtering.

## 3.2 Sputtering of $CO_2$ and other species from hydrocarbons with different functional groups

Figure 3 compares the $CO_2$ sputtering of $H_2O$:hexane (60:40), $H_2O$:hexanol (99:1) and $H_2O$:hexanoic acid (99:1) ice. (Unfortunately, the 60:40 $H_2O$:hexane ratio exhibited poor S/N, so a 99:1 mixture of $H_2O$:hexane experiment was not undertaken.) The COOH unit in hexanoic acid appears to enable $CO_2$ to be generated more quickly and in larger abundances than with the other functionalized organics. With hexanol, and especially with hexane, oxidation of the material is slower than in hexanoic acid (Figure 3). Since one of the carbon-oxygen bonds are already present in the hexanol molecule, we hypothesize that the presence of this functional group helps to accelerate the production of $CO_2$ relative to its production in the hexane ices.

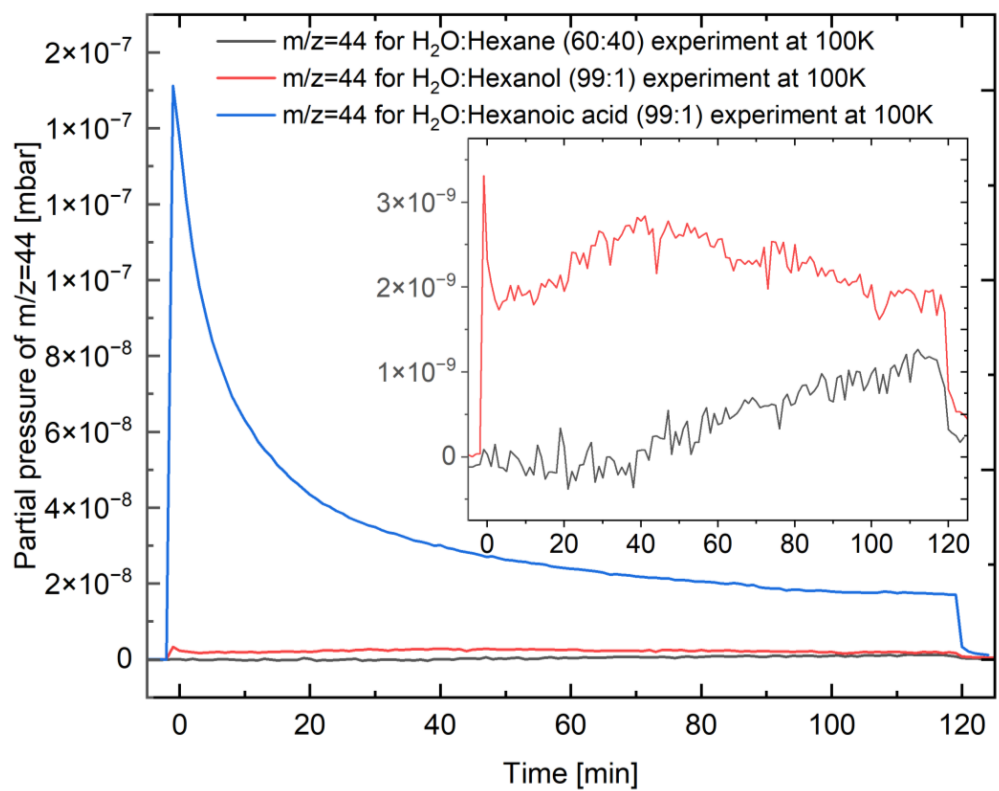

Figure 3: Background-subtracted sputtered evolution of $CO_2$ (m/z=44) for $H_2O$:hexane (60:40), $H_2O$:hexanol (99:1) and $H_2O$:hexanoic acid (99:1) at 100 K. Inset shows the sputtering of $CO_2$ for $H_2O$:hexane (60:40) and $H_2O$:hexanol (99:1) ice. The initial jump in $CO_2$ production is ~100x higher in hexanoic acid than hexane or hexanol samples, suggesting that the -COOH functional group is responsible for rapid and efficient $CO_2$ formation.

Figure 4 compares the sputtered evolution of a simple hydrocarbon fragment m/z=CH$_3$ from H$_2$O:hexane (60:40), H$_2$O:hexanol (99:1), and H$_2$O:hexanoic acid (99:1) ices at 120 K. The evolution of m/z=15 shows different profiles for the differently functionalized hydrocarbons. In case of H$_2$O:hexane ice, the intensity of m/z=15 increases more slowly than in H$_2$O:hexanol and H$_2$O:hexanoic acid ices. As with CO$_2$, m/z=15 also appears at a significantly higher intensity for the acid-functionalized compared to the alcohol-functionalized and non-functionalized hexane. This may be due to the ejection of pentane after the COOH group is lost (see Section 3.4.3 for details).

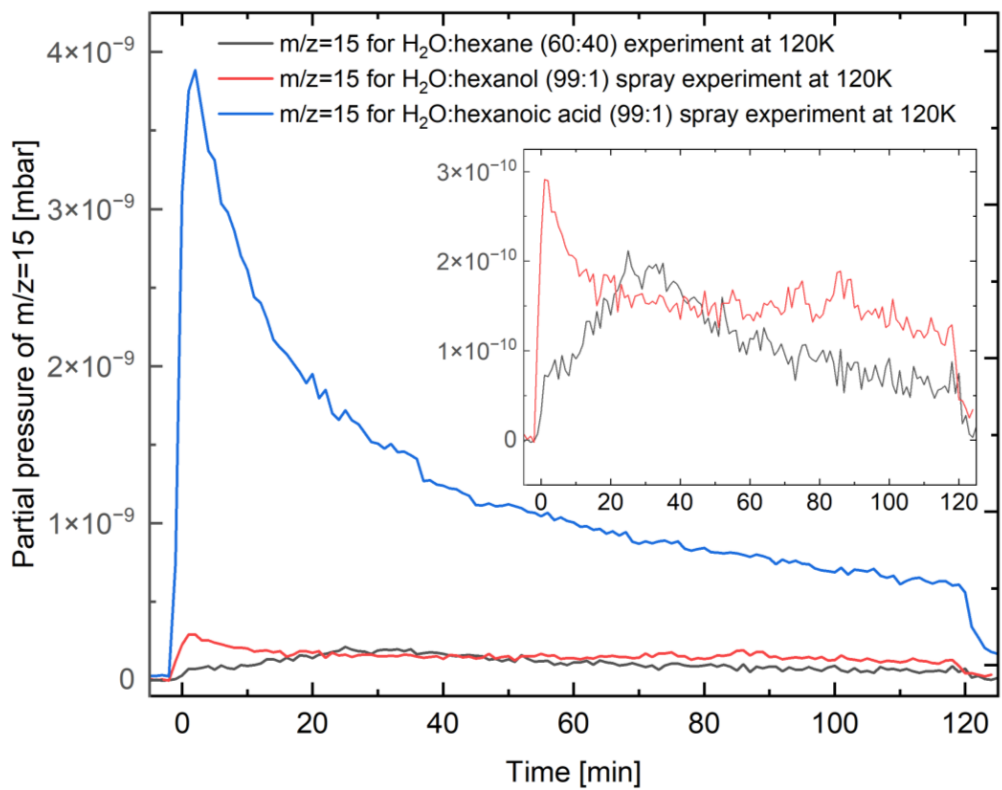

Figure 4: Background-subtracted sputtered evolution of m/z=15 for H$_2$O:hexane (60:40), H$_2$O:hexanol (99:1) and H$_2$O:hexanoic acid (99:1) at 120 K. Inset shows the sputtering of m/z=15 for H$_2$O:hexane (60:40) and H$_2$O:hexanol (99:1) ice. m/z=15 from H$_2$O:hexanoic acid, when integrated, is over 10x higher than m/z=15 from H$_2$O:hexane and H$_2$O:hexanol ice sputtering. Some of this 15 m/z can be attributed to a fragment of pentane, which is formed when the COOH group is lost from the hexanoic acid (see Section 3.4.3 for details). Other hydrocarbon fragments are shown in Figs. A5 and A6.

### 3.3 Temperature dependence of CO$_2$ sputtering

Sputtering yields are known to change with temperature (Davis et al., 2021) and in our prior work, we found that sputtering from H$_2$O:hexane ices increased from 80 K to 100 K to 120 K. In Figure 5a, the evolution of CO$_2$ for each of the functional group types is shown for three different temperatures. The sputtered CO$_2$ for each experiment is subtracted by the background CO$_2$ in pure $H_2O$ ice. The background-corrected sputtered (integrated) $CO_2$ was found to be over 10x times higher for hexanoic acid at 100 K and 120 K than in the ices with the hexanol. The sputtered $CO_2$ from $H_2O$:hexanoic acid ice was 3x times higher at 100 K and 120 K compared to 80 K.

To understand the lower $CO_2$ sputtering at 80 K, a temperature programmed desorption (TPD) experiment was performed for each sputtering experiment and is shown in Figure 5b. In these experiments, the temperature was ramped to 140 K following sputtering at each of the three temperatures, with a ramping rate of 1 K/min in all cases and the spectra were collected every minute. During the TPD experiment, trapped $CO_2$ outgassed from the respective irradiated ices. The TPD profile (black dashed) shows outgassing of $CO_2$ that is produced by radiolysis of hexanol and $H_2O$.

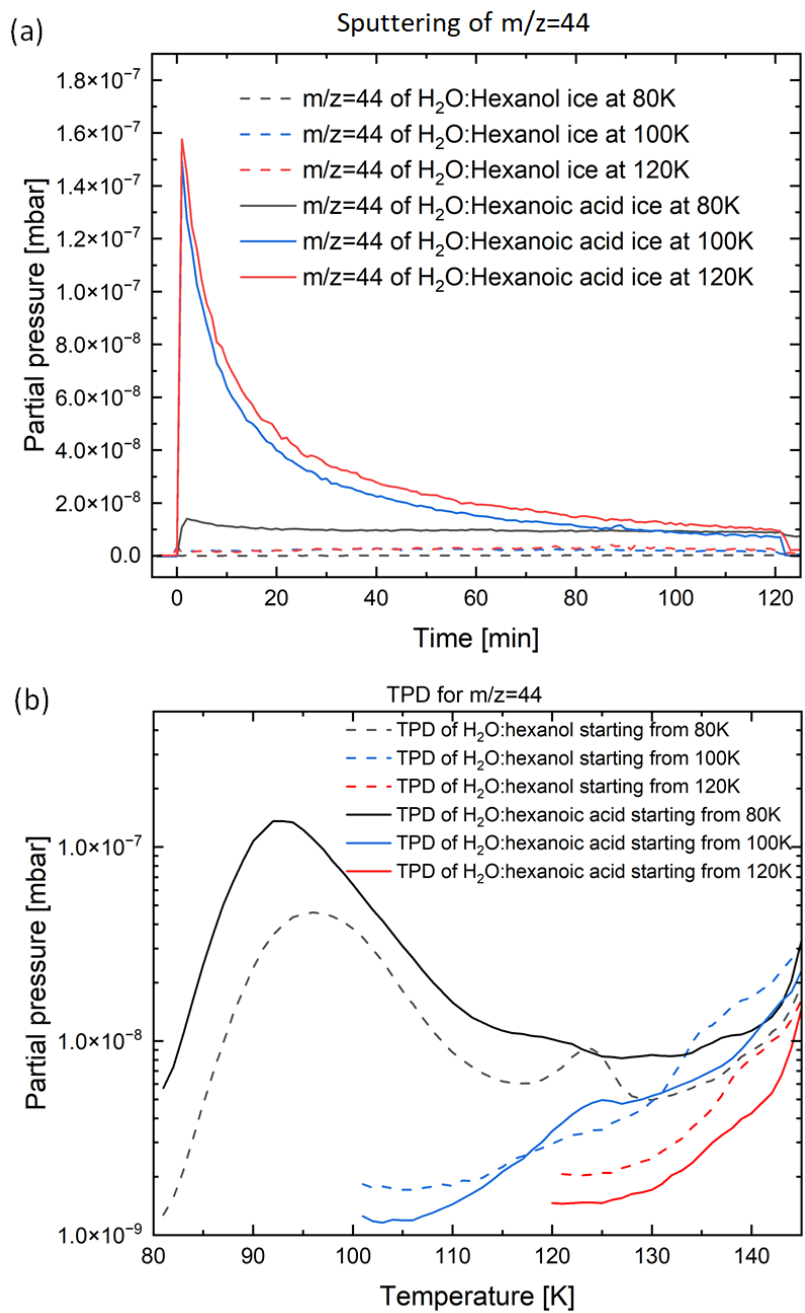

Figure 5: (a) Background-subtracted $CO_2$ evolution during sputtering of $H_2O$:hexanol (99:1) (shown as dashed lines) and $H_2O$:hexanoic acid (99:1) (shown as solid lines) ice at 80 K (black), 100 K (blue), and 120 K (red). $CO_2$ sputtering was over 10x higher from $H_2O$:hexanoic acid than from $H_2O$:hexanol at every temperature studied. (b) $CO_2$ temperature-programmed desorption from irradiated $H_2O$:hexanoic acid (shown as solid lines) and $H_2O$:hexanol (shown as dashed lines) ices at 80 K, 100 K and 120 K, showing that $CO_2$ was generated and trapped during the 80 K irradiation and was released at ~90 K during warming of the sample ice.

The integrated sputtering intensity, TPD intensity and the sum of sputtering and the TPD amounts are shown in Figure 6 for $H_2O$:hexanoic acid and $H_2O$:hexanol ice. The total amount of $CO_2$ generated was ~75-85% less in the hexanol ice than in the hexanoic acid ice at 80 K, 100 K and 120 K. In both ices, $CO_2$ appears to be sputtered less (and trapped more) at 80 K than at the higher temperatures. For hexanoic acid, we find that about 36% of the total $CO_2$ generated in the 80 K experiment was sputtered vs. 97% sputtered in the 120 K experiment. Because of high pressures in the chamber, we were unable to collect mass spectra past 145 K. It is possible that some $CO_2$ is still trapped in ice at this temperature (Gudipati et al., 2023), and our integrated TPD data values may be underestimating the $CO_2$ that is trapped at each temperature.

For hexanoic acid, the combined sputtering and TPD amounts (top panel of Fig. 6) suggest that $CO_2$ was formed in similar amounts at all studied temperatures. $CO_2$ formation in the hexanol experiments, however, was not constant – it dropped by over 40% at 100 and 120 K (bottom panel of Fig. 6).

Taken together, these results suggest that the $CO_2$ production mechanism in $H_2O$:hexanol is different than the $CO_2$ production mechanism in $H_2O$:hexanoic acid ice. This is likely due to the presence of COOH in $H_2O$:hexanoic acid, which can generate $CO_2$ by simple bond cleavage, whereas hexanol requires further oxidation. As the hexanol is oxidized, it is possible that volatile intermediate fragments could be lost to the atmosphere, reducing the amount of $CO_2$ that can be produced during our 120-minute experiments.

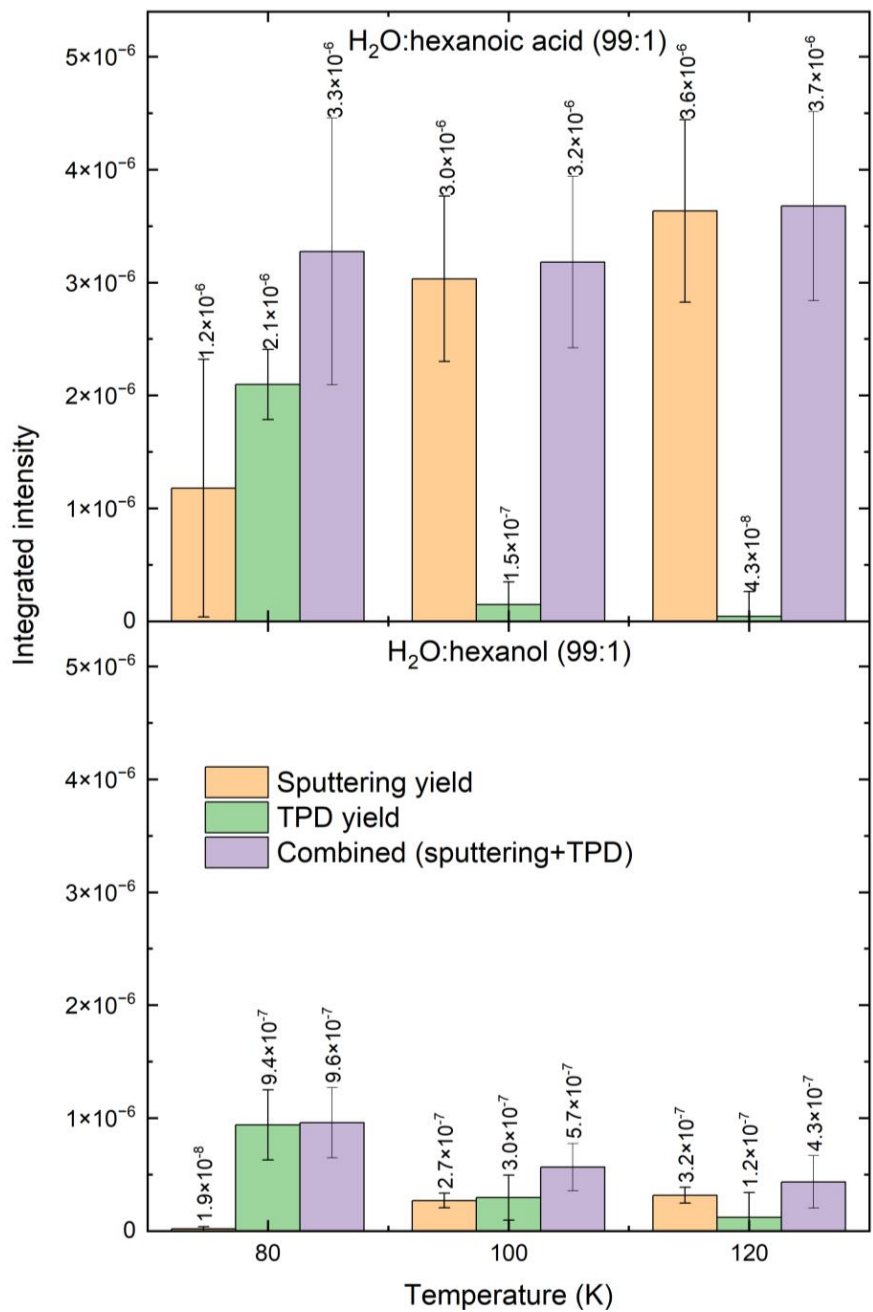

Figure 6: Integrated sputtering yield, TPD yield and combined sputtering and TPD yield of $CO_2$ are shown for $H_2O$:hexanoic acid (99:1) ice (top) and for $H_2O$:hexanol (99:1) ice (bottom) at three different temperatures. Multiple $H_2O$:hexanoic acid sputtering experiments (n≥3) were conducted for each temperature, and the given values represent the averages with error bars representing one standard deviation. Relative standard deviations for sputtered $H_2O$:hexanol ices were assumed to be the same as those of $H_2O$:hexanoic acid. The error bars for all TPD experiments were based on relative standard deviations calculated from prior TPD experiments using this system (n≥3). Sputtering of $CO_2$ increases with temperature, but TPD yield of $CO_2$ is highest at 80 K, indicating that it can be produced and remain trapped at 80 K.

## 3.4. Temperature dependence of hydrocarbon sputtering

### 3.4.1 Mass spectrometer detection of small chain hydrocarbons

Temperature-dependent behaviors of different hydrocarbon fragments are shown in Figure A5 and A6 for $H_2O$:hexanol and $H_2O$:hexanoic acid ices, respectively. For this analysis, we chose the fragmented hydrocarbons m/z=15, m/z=27, m/z=39 and m/z=41, which correspond to $CH_3$, $C_2H_3$, $C_3H_3$, and $C_3H_5$. These particular fragments are chosen to avoid any potential confusion from the oxygen-incorporated species (e.g., avoiding $CH_4$ because of overlap with oxygen at m/z 16). In our prior work with hexane:$H_2O$ ices, we noted that these sputtered hydrocarbon components increased for the first ~30 minutes of sputtering and then decreased. In hexanol and hexanoic acid, the intensity of all these hydrocarbon fragments have a faster rise (reaching a maximum within 10 minutes), followed by a decay. The sputtering profile of these hydrocarbon fragments are similar, and it is possible that they originate from a parent ion that is fragmented in the quadrupole mass spectrometer. All of these fragments also follow a temperature-dependent trend – as the temperature rises, the sputtered intensity of hydrocarbon fragments increases. In the case of $H_2O$:hexanoic acid, the sputtered intensities of the fragmented hydrocarbons are significantly higher (an increase of ~80%) for 120 K compared to 80 K and 100 K (see Figure A6).

### 3.4.2 Sputtering of long chain hydrocarbons

The sputtering intensities of the long chain hydrocarbons pentane (m/z=72) and hexane (m/z=86) are plotted in Fig. 7, along with nearby masses such as m/z=71, 73 and 74. Pentane and hexane can be formed by loss of -COOH and -OH groups from the hexanoic acid and hexanol molecules, respectively, followed by hydrogen atom abstraction from the hydrogen-rich ice, and their assignments were confirmed by temperature-programmed desorption (TPD) experiments (see Section 3.4.3).

Figure 7a and 7b shows the sputtering of these masses for $H_2O$:hexanol and $H_2O$:hexanoic acid ice at 120 K temperature. Sputtering of m/z=72 was found to occur in both samples, but ~100x more pentane was sputtered from $H_2O$:hexanoic acid than from $H_2O$:hexanol ice (integrated traces from Fig. 7a and 7b). Sputtering of m/z=86 (hexane) was much less efficient, with the integrated m/z=86 intensity amounting to only 15% and 1% of the m/z=72 intensity for $H_2O$:hexanol and $H_2O$:hexanoic acid respectively. TPD results (described below) suggest that this difference is due to the higher volatility of pentane vs. hexane.

### 3.4.3 Identification of butane, pentane, and hexane from TPD profiles of $H_2O$:hexanol and $H_2O$:hexanoic acid ice

TPD experiments were conducted to determine whether specific species could remain trapped in the ice following irradiation in our experiments. Following sputtering at 80 K, the irradiated ices were ramped at 1 K/min to 145 K (where the pressure becomes too high to continue collecting mass spectral data). The TPD profiles are shown in Figure 7c and 7d for $H_2O$:hexanol and $H_2O$:hexanoic acid, respectively. The TPD of both $H_2O$:hexanol and $H_2O$:hexanoic acid ice systems give a peak at 95 K that is associated with masses m/z=15, m/z=27 and m/z=43. After comparing the sublimation temperature with the literature (butane sublimation occurs at ~90-95 K) (Abplanalp and Kaiser, 2016; Jones and Kaiser, 2013) and after comparing these masses with the butane mass fragmentation spectrum from NIST (Figure 7e) (E. Wallace, n.d.), we assigned the TPD peak at 95 K to butane. In case of $H_2O$:hexanoic acid, a TPD peak at 120 K appears for several masses including m/z=15, 27, 39, 41, 43, 57 and 72. Here, the intensity ratios of the fragments fit well with the NIST electron ionization mass spectrum of pentane (E. Wallace, n.d.), which is shown in the pink in panel 7e. Based on this and prior desorption studies (Abplanalp and Kaiser, 2016; Jones and Kaiser, 2013), we assigned the 120 K peak to pentane.

In case of $H_2O$:hexanol ice (Panel 7c) another TPD peak is observed at 125 K that is not associated with pentane. NIST data shows that the hexane fragments primarily to m/z=57, 43, 41, 29, and 27 (E. Wallace, n.d.), which are all present at 125 K. Based on this and prior desorption studies (Abplanalp and Kaiser, 2016; Jones and Kaiser, 2013), we assign the 125 K TPD peak in the $H_2O$:hexanol ice system to hexane.

These TPD experiments suggest that hexanol bond cleavage can occur at the C-OH bond to produce hexane (nonvolatile at 120 K), whereas in hexanoic acid the C-COOH bond cleavage produces pentane (volatile at ≥120 K) and $CO_2$ (volatile at ≥100 K). At 120 K, hexane remains on the surface while pentane can be sputtered, and so hexanol generates fewer sputtered hydrocarbon signatures than hexanoic acid (also compare Figures A5 and A6). These experiments suggest that the radiation-induced chemistry of both an icy body's surface and its exosphere may be critically dependent on the bonds that break, on the volatilities of the resulting species, and ultimately on the surface temperature which controls whether the surface species are lost to the exosphere.

Our experiment suggests that acids generate higher amounts of $CO_2$ and could leave behind shorter carbon chains. On Earth, fatty acids (hydrocarbons with carboxylic acid functional groups similar to hexanoic acid) play an important role in forming cell membranes and are thought to be important diagnostic biosignatures for detecting life in space (Georgiou and Deamer, 2014). Our work suggests that fatty acids, if present on irradiated icy surfaces, may sputter $CO_2$ and leave behind hydrocarbon chains. Depending on local temperatures, volatile hydrocarbon chains and other fragments could be sputtered to the atmosphere, and less volatile ones could remain on the surface. In either case, once the carboxylic acid functional group is lost, a fatty acid would be no longer identifiable as a distinct biosignature, either on the surface or in the exosphere.

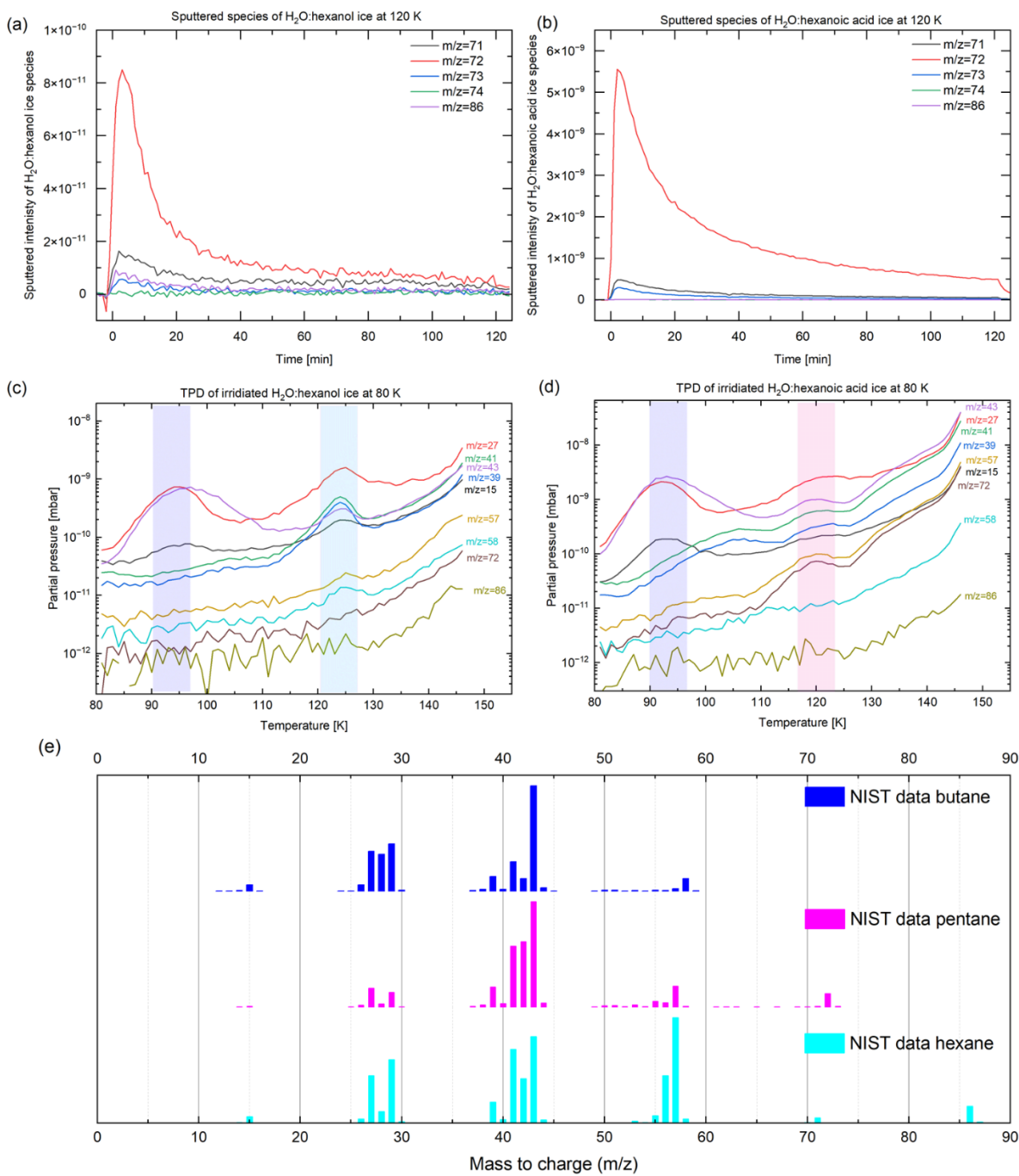

Figure 7: Sputtered profiles of m/z=71, m/z=72, m/z=73, m/z=74 and m/z=86 are shown for (a) $H_2O$:hexanol and (b) $H_2O$:hexanoic acid ice. Integrated intensity of m/z=72 from $H_2O$:hexanoic acid is 100x higher than m/z=72 from $H_2O$:hexanol ice suggesting that, once the carboxylic acid is cleaved, pentane is sputtered from $H_2O$:hexanoic acid ice. TPD profiles of different masses are shown for (c) $H_2O$:hexanol and (d) $H_2O$:hexanoic acid ice. (e) NIST data for mass fragmentation of butane, pentane and hexane. The peaks at 95 K (dark blue) were assigned to butane fragment masses generated in the mass spectrometer. The peaks at 120 K (pink) were assigned to pentane fragments and peaks at 125 K (light blue) to hexane fragments.

## 4. Conclusion

The sputtering of organics is critical for understanding the relation between exospheric and surface composition on icy radiation-drenched bodies. When exposed to 2 keV electrons, our results show that organics can generate $CO_2$ in different amounts depending on the types of substituents. Hydrocarbons containing COOH units can generate $CO_2$ more quickly compared to those with OH substituents. For a fresh deposit on Europa's surface (for example, from a plume enriched with organics having OH and COOH functional group), we expect organics containing -COOH functional groups would sputter $CO_2$ more efficiently than organics with OH functional groups, possibly leading to local $CO_2$ enhancements. It is currently not clear what concentrations and types of organics would be required to generate a local $CO_2$ enhancement significant enough to be observed by spacecraft instrumentation. It should be noted that $CO_2$ is a very common molecule in space and more studies are needed to determine whether it could be generated from irradiation of non-organic sources such as carbonate minerals.

The key findings from this experimental work are as follows:

- The integrated increase in $CO_2$ sputtering with electron radiation was over 10x times higher for hydrocarbon containing a COOH group than for our hydrocarbons containing only an OH group or with no oxygen-containing group. This suggests that biosignatures containing acid-functionalized organics (e.g., fatty acids) may be more prone than other organics to sputtering of $CO_2$. Still, it is possible for other hydrocarbons to eventually generate $CO_2$ with continued radiation exposure.
- The overall amount of $CO_2$ sputtering is temperature dependent: from hexanoic acid it is 3x lower at 80 K than at 120 K or 100 K. With hexanol, it is 15x lower at 80 K than it is at 120 or 100 K. Because of its sublimation point near 80 K, we expect that $CO_2$ remains trapped at Europa (or other icy bodies) in areas where the temperature is ≤80 K (i.e., near the poles), whereas in the higher-temperature equator or warmer, illuminated dayside areas it can be sputtered directly into the exosphere.
- We found that short hydrocarbons can be sputtered into the exosphere with keV electrons. In the case of Europa, sputtering could release a maximum of up to five-carbon hydrocarbon chains into the exosphere, as the maximum temperature of Europa is close to 120 K (Ashkenazy, 2019). Following loss of a COOH group, small organic fragments could become volatile enough to escape. The combined effects of radiolysis and desorption may make it difficult to retain (and detect) small organics and their radiation byproducts on highly irradiated icy surfaces.

$CO_2$ is detected in most icy worlds (Carlson, 1999; Combe et al., 2019; Hibbitts et al., 2003; Nixon, 2024) and is a common byproduct of radiation. Though hydrocarbons have not yet been found on Europa, they have been observed in Enceladus (Waite et al., 2006), Callisto and Ganymede (McCord et al., 1997) and many types of organics with varying functional groups have been detected in comets (Hänni et al., 2022). Our study shows that radiolysis generates different amounts of sputtered $CO_2$ depending on which functional groups are present. In addition, these processes depend significantly on temperature: if local temperatures are ≤80 K, sputtering of $CO_2$ into the exosphere is less efficient and may be able to build until sputtered or desorbed at warmer temperatures (e.g. during the daytime or a warmer season). Future spacecraft flybys may be able to identify sputtered species such as small hydrocarbons and $CO_2$ from deposits on the surface. However, we note that radiation-sputtered $CO_2$ could also arise from non-organic sources, such as carbonate salts or seawater, and that more work is needed to understand the sputtering of these types of analogs under similar conditions.

## Acknowledgement

This research was carried out at the Jet Propulsion Laboratory, California Institute of Technology, under a contract with the National Aeronautics and Space Administration. S.C., B.H. and M.S.G. acknowledge support from NASA's Habitable Worlds and Solar System Workings Programs.

## Copyright statement

# Appendix

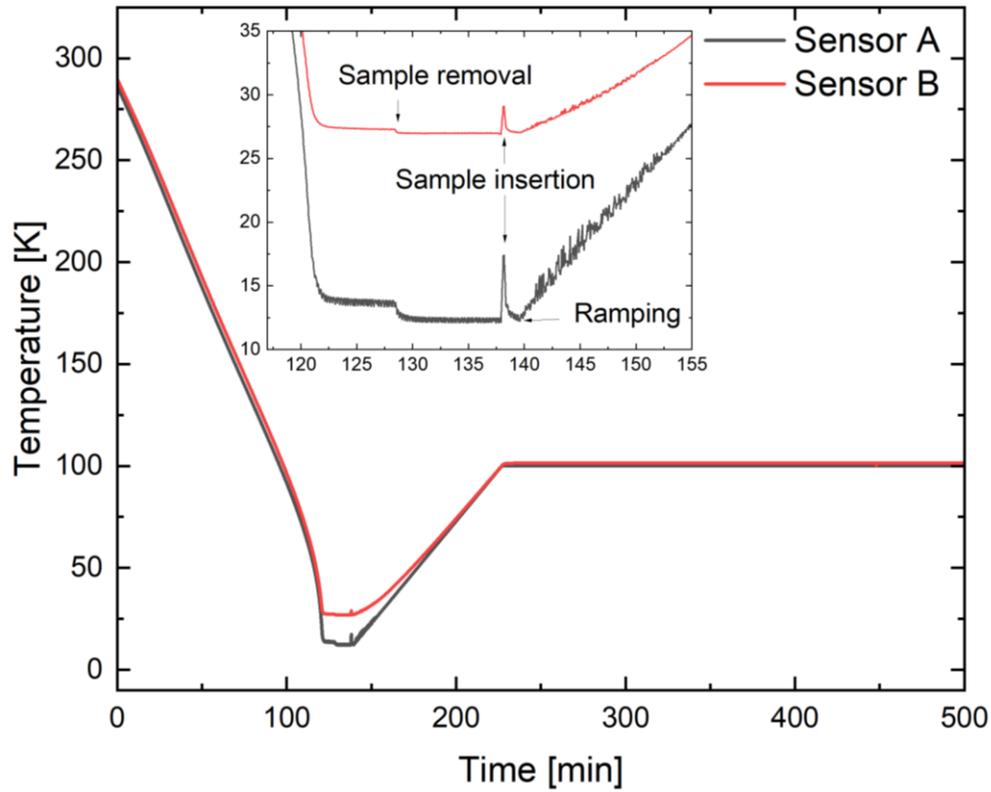

Figure A1: Temperature variation of two sensors positioned in two different places of the sample holder during the course of the full sputtering experiments starting from sample preparation. Sensor A is positioned at the bottom of the cold head and Sensor B is placed close to the sample holder. A temperature variation of 10 K is observed at temperatures in the 30 K to 45 K range. However, during the sputtering experiments, for example at 80 K, 100 K and 120 K, both diodes measured the same temperature.

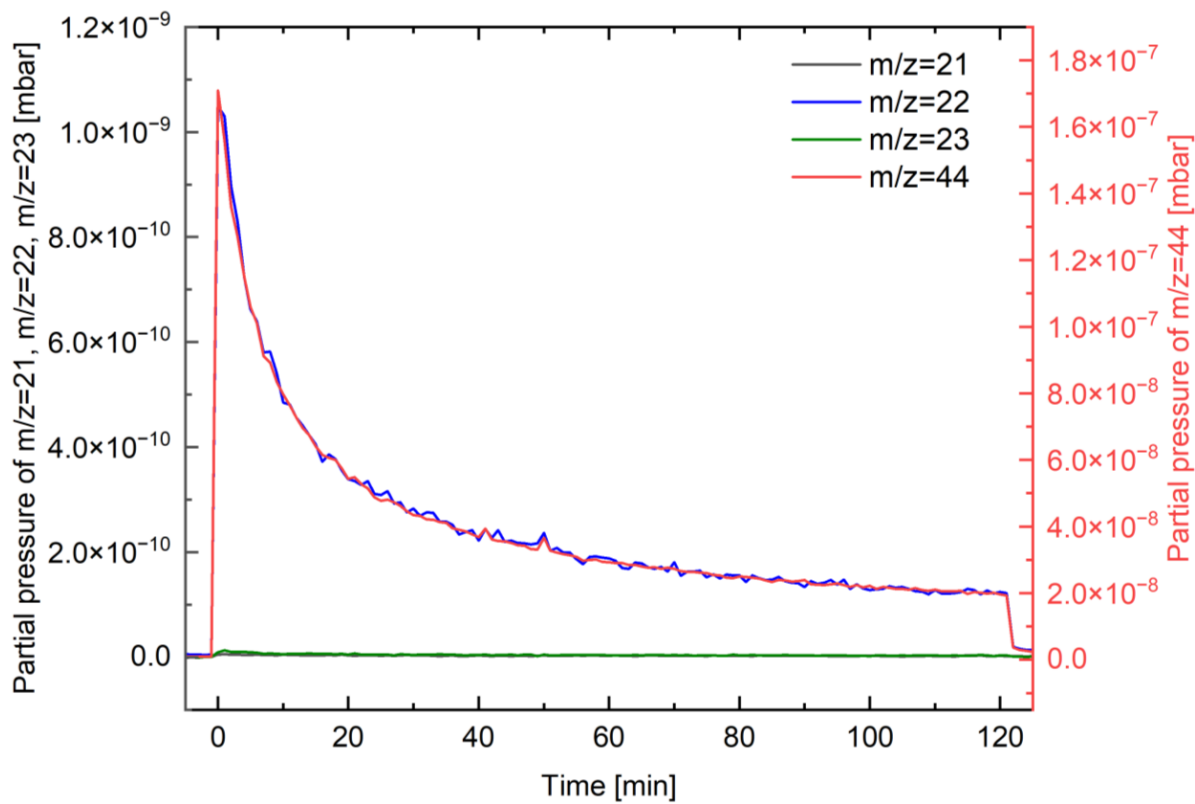

Figure A2: Sputtered evolution of mass to charge ratio 21, 22, 23 and 44 are shown for the $H_2O$:hexanoic acid experiment at 120 K. m/z=22 corresponds solely to doubly-charged $CO_2$, which is ~1% of the m/z=44, which is comparable to NIST $CO_2$ mass spectral data (no other organic molecule exists with a mass of m/z 22). The sputtered intensity of m/z=22 is further compared with the nearby background masses of m/z=21 and 23, which are both much lower than the m/z=22. The evolution of the blue and red traces for m/z=22 and 44 overlap, suggesting that the majority of m/z 44 can also be directly attributed to $CO_2$ and not any other organic fragment.

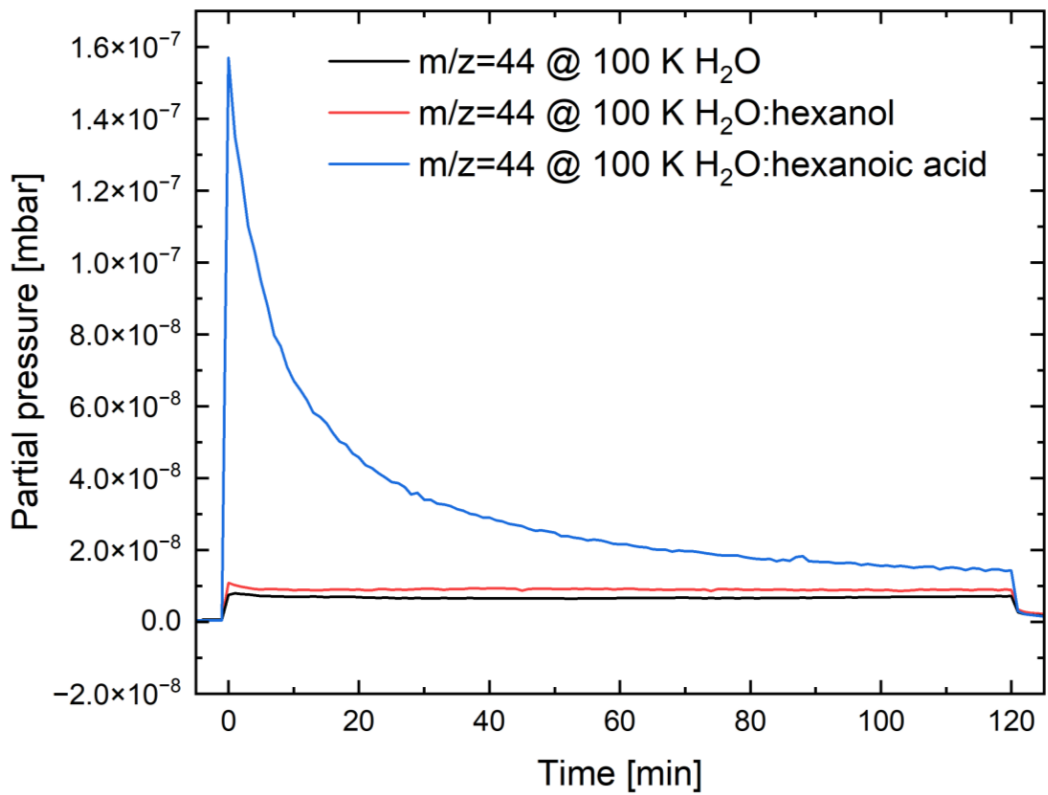

Figure A3: Sputtering of $CO_2$ (m/z=44) is shown for pure $H_2O$ ice, $H_2O$:hexanol (99:1) ice and $H_2O$:hexanoic acid (99:1) ice at 100 K. The total $CO_2$ sputtered from the pure water control experiment is 22% of the total $CO_2$ produced from $H_2O$:hexanoic acid ice and 75% of the total $CO_2$ produced from $H_2O$:hexanol ice. This background $CO_2$ contribution is subtracted for each experiment.

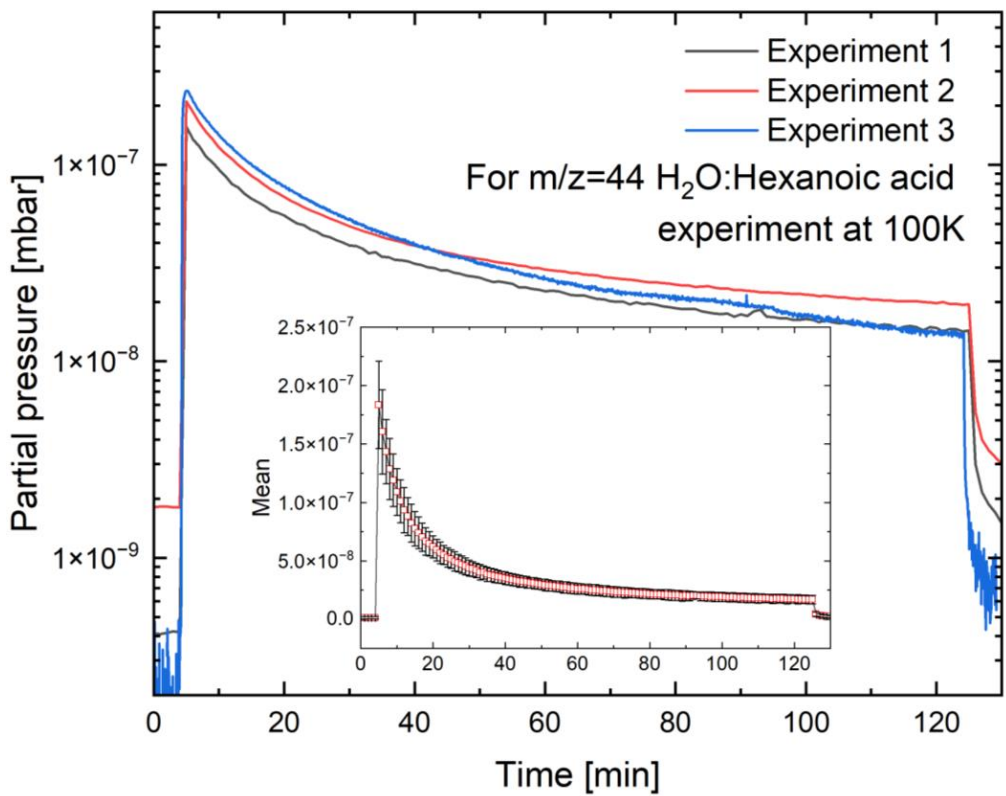

Figure A4: Comparison of $CO_2$ sputtering from three sets of $H_2O$:hexanoic acid ice experiments at 100 K. Inset shows the average and error bars representing one standard deviation of the three experiments. The relative standard deviation is ~30% for the first 10 minutes and then it reduces to ~3% afterward.

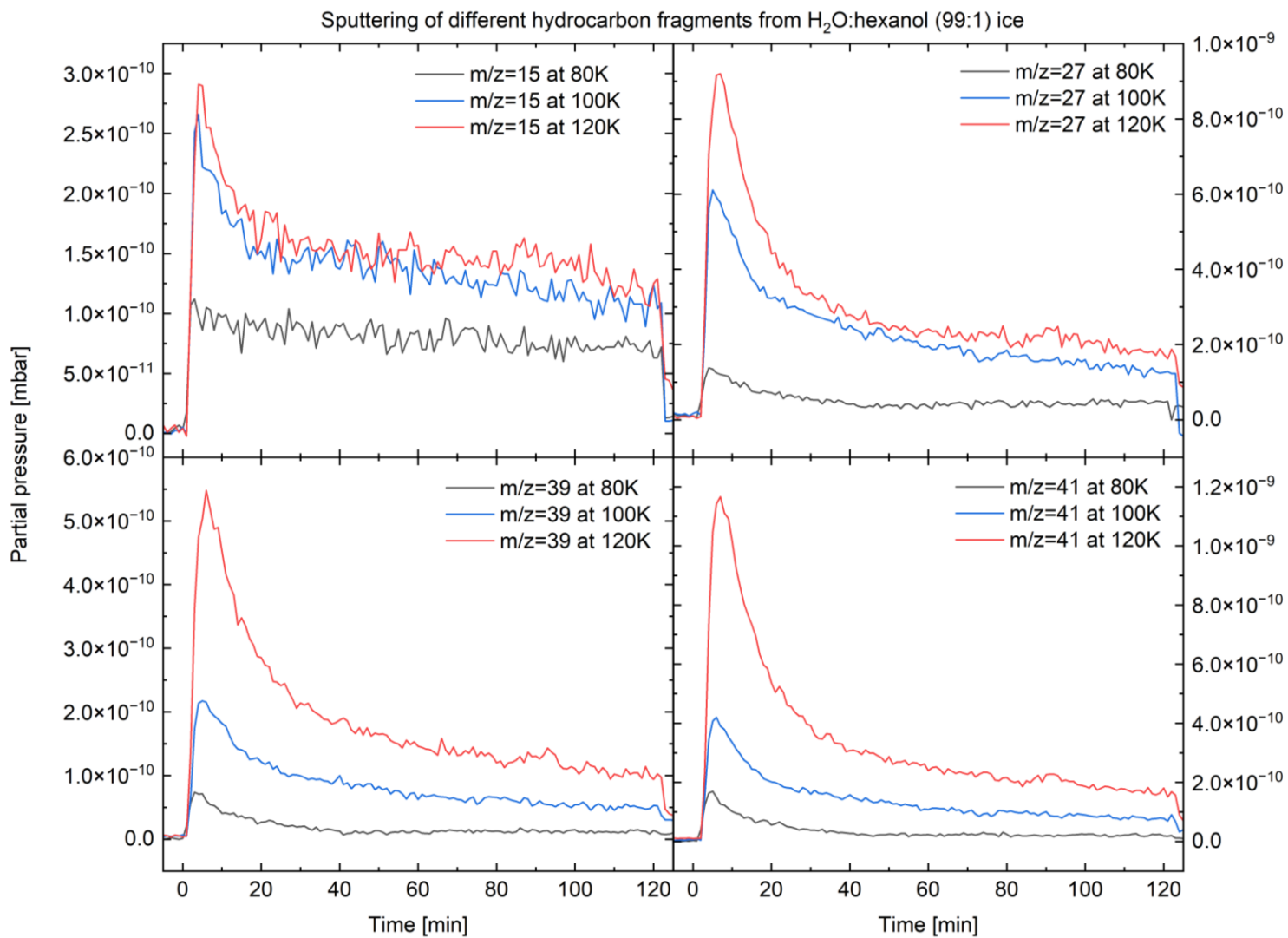

Figure A5: Background-subtracted sputtered evolution of fragmented hydrocarbons for m/z of 15 ($CH_3$), 27 ($C_2H_3$), 39 ($C_3H_3$) and 41 ($C_3H_5$) are depicted for three different temperatures for the $H_2O$:hexanol experiment. Each curve has a similar shape, with more hydrocarbon sputtering occurring as the temperatures are increased.

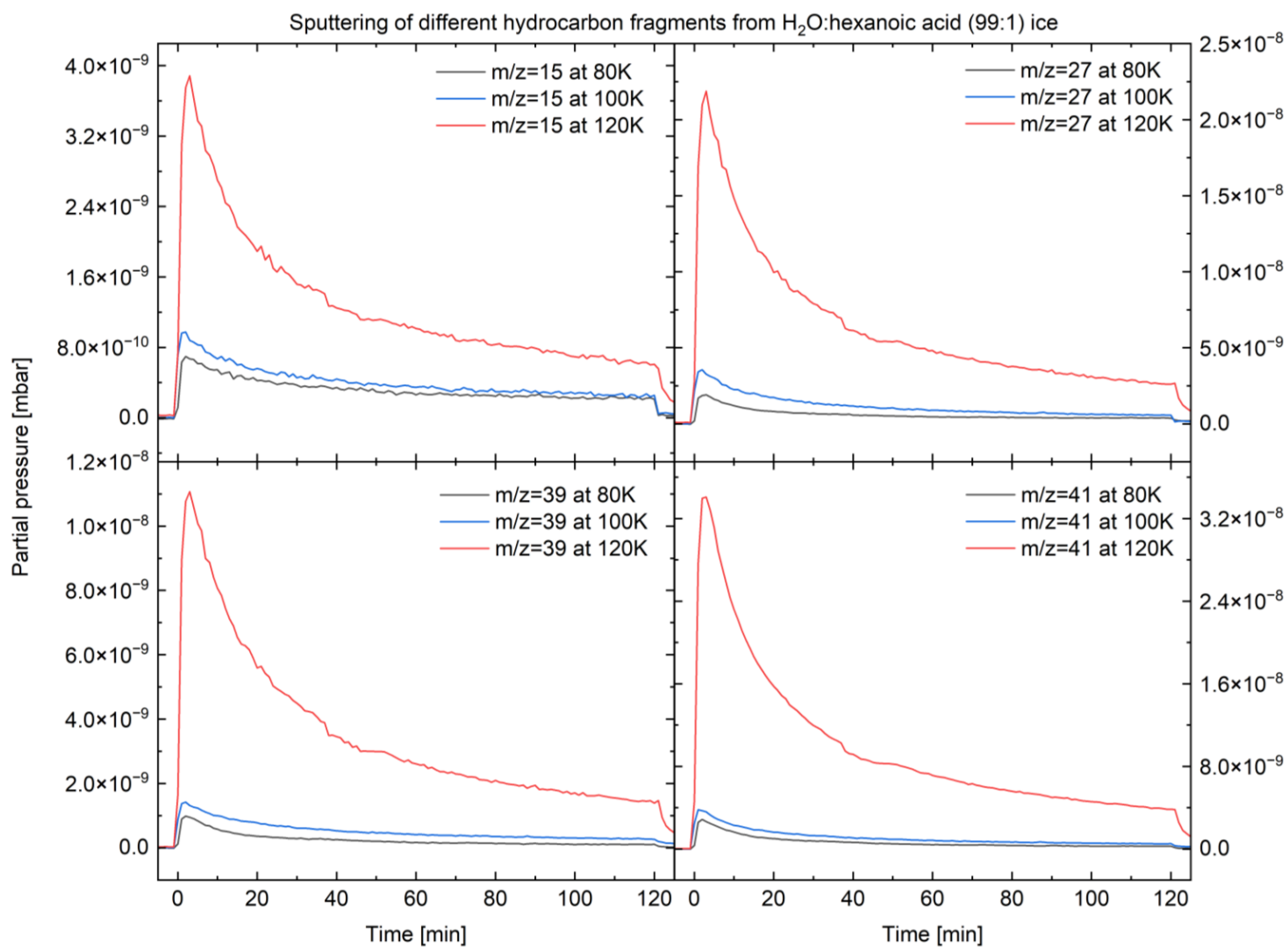

Figure A6: Background-subtracted sputtered evolution of fragmented hydrocarbons for m/z of 15 ($CH_3$), 27 ($C_2H_3$), 39 ($C_3H_3$) and 41 ($C_3H_5$) are depicted for three different temperatures for the $H_2O$:hexanoic acid experiment. Each curve has a similar shape, with more hydrocarbon sputtering occurring as the temperatures are increased, particularly when the experiment is conducted at 120 K.